\documentstyle[12pt]{article}

\newcommand{\EQ}{\begin{equation}}
\newcommand{\EN}{\end{equation}}

\begin{document}

\topmargin 0pt
\oddsidemargin 5mm
\newcommand{\NP}[1]{Nucl.\ Phys.\ {\bf #1}}
\newcommand{\PL}[1]{Phys.\ Lett.\ {\bf #1}}
\newcommand{\NC}[1]{Nuovo Cimento {\bf #1}}
\newcommand{\CMP}[1]{Comm.\ Math.\ Phys.\ {\bf #1}}
\newcommand{\PR}[1]{Phys.\ Rev.\ {\bf #1}}
\newcommand{\PRL}[1]{Phys.\ Rev.\ Lett.\ {\bf #1}}
\newcommand{\MPL}[1]{Mod.\ Phys.\ Lett.\ {\bf #1}}
\newcommand{\JETP}[1]{Sov.\ Phys.\ JETP {\bf #1}}
\newcommand{\TMP}[1]{Teor.\ Mat.\ Fiz.\ {\bf #1}}

\renewcommand{\thefootnote}{\fnsymbol{footnote}}

\newpage
\setcounter{page}{0}
\begin{titlepage}
\begin{flushright}
SISSA-91-EP-168
\end{flushright}
\vspace{0.5cm}
\begin{center}
{\large  The thermodynamic Bethe ansatz for deformed $WA_{N-1}$ conformal field
theories} \\
\vspace{1cm}
\vspace{1cm}
{\large  M\'arcio Jos\'e  Martins
\footnote{on leave from Departamento de Fisica, Universidade Federal de
S.Carlos, C.P. 676 - S.Carlos 13560, Brazil}
\footnote{martins@itssissa.bitnet}} \\
\vspace{1cm}
{\em International School for Advanced Studies \\ Strada Costiera 11\\
34014, Trieste, Italy } \\
\end{center}
\vspace{1.2cm}

\begin{abstract}
We propose and investigate the thermodynamic Bethe ansatz equations for
the minimal $W_p^N$ models~(associated with the $A_{N-1}$ Lie algebra)
perturbed by the least~($Z_N$ invariant) primary field $\Phi_N$.
Our results reproduce the expected ultraviolet and infrared regimes. In
particular for the positive sign of the perturbation our equations
describe the behaviour of the ground state flowing from the $W_p^N$ model to
the next $W_{p-1}^N$ fixed point.
\end{abstract}
\vspace{.5cm}
\centerline{October 1991,~Phys.Lett.B,~in press}
\vspace{.3cm}
\end{titlepage}

\renewcommand{\thefootnote}{\arabic{footnote}}
\setcounter{footnote}{0}

\newpage
Recently Al.Zamolodchikov has pointed out the
importance of the thermodynamic
Bethe ansatz~(TBA) in the context
of the non-diagonal scattering theories~[1,2].
In particular, for the RSOS models,
the respective TBA equations for the
ground state reproduce the expected
ultraviolet and infrared behaviours. The
RSOS scattering theory describe~[1,3]
the scaling behaviour of the minimal
conformal $M_p$ field theories perturbed
by the $\phi_{1,3}$ operator. For the
negative sign of the perturbation,
the TBA equations describe the massive
behaviour of the (p-1) $Z_2$ symmetric kinks.
On the other hand, for the
positive sign, they describe the
Renormalization Group~(RG) trajectory flowing
from the $M_p$ critical point
to the $M_{p-1}$ fixed point.
This is a well known example of
RG flow between two non-trivial
fixed points~[4,5]. It seems important to
generalize this approach to other
systems possessing ``higher'' symmetries. The
natural candidates are the $Z_N$-invariant
conformal field theories related to
the minimal $W_p^N$ models~[6] with central charge
$c=(N-1)(1-\frac{N(N+1)}{p(p+1)})$,
$p=N+1,N+2,...$ . The analogue of the
$\phi_{1,3}$ operator in the $W_p^N$
theories is the least~($Z_N$ invariant)
 relevant primary
field $\Phi_N$ associated with the
weight of the adjoint representation of
$SU(N)_{p+1-N}$, which has conformal dimension
$\Delta_{\Phi_N}=1-\frac{N}{p+1}$.
For the sake of simplicity we first
consider the $W_p^3$ theory. The perturbed
action is defined by
\begin{eqnarray}
A_{W_p^3}= A_{CFT} + \lambda \int d^2 x \
phi \left [ \begin{array}{cc}
1 & 2 \\ 1 & 2 \\ \end{array} \right ]
\end{eqnarray}
where the field $\Phi_3$ is identified with the operator
 $\phi \left [ \begin{array}{cc}
1 & 2 \\ 1 & 2 \\ \end{array} \right ]$
in the notation of ref.[6].

 The field $\phi \left [ \begin{array}{cc}
1 & 2 \\ 1 & 2 \\ \end{array} \right ]$
preserves the global symmetry of
the $W_p^3$ model and therefore the
perturbed action (1) is
still $Z_3$ invariant.
Moreover, the action (1) defines
an integrable theory~[7,8,9]
and its physical behaviour
depends on the sign of the coupling constant $\lambda$.
For $\lambda<0$ the
action (1) describes the scaling region
of a generalized $A_{N-1}$-RSOS~[10]
model, and its kink-kink S-matrices have
been proposed in ref.[11] .
In the case of $\lambda>0$, based on
the product expansion properties of
the field $\phi \left [ \begin{array}{cc}
1 & 2 \\ 1 & 2 \\ \end{array} \right ]$
and the RG analysis~[8,12],
one expects that the theory $W_p^3$
flows to the $W_{p-1}^3$ fixed
 point~[8,12,13], $p>4$. We will
 henceforth refer to these theories
as $W_p^{3(\pm)}$ models. Up to now
there has not been any general
scheme in order to derive the TBA
equations directly from the non-diagonal
S-matrices. However, Al. Zamolodchikov
has pointed out
that~(see also refs.[20],[21])
in the usual RSOS-models~[1,2,19],
the final result assumes a general
form in terms
of coupled integrals equations with
a universal kernel, respecting the
symmetries of the kink interactions.
Then, using the available
symmetries and the
knowledge of certain known limits it
is not difficult to guess the associated
TBA equations. The validity of
these equations can be confirmed
through checks done a posteriori.
For example, analyzing the ultraviolet
and infrared behaviour of
the proposed TBA equations and also comparing them
with the respective perturbation theory.
The TBA equations that we propose for the action (1)
are the following
\EQ
\epsilon_i^a(\theta) -\nu_i^a(\theta)
+\frac{1}{2 \pi} \sum_{j=1}^{p-3} l_{i,j}
\sum_{b=1}^{2} \int_{- \infty}^{\infty}d
\theta' K_{a,b}(\theta-\theta')
Ln(1+e^{-\epsilon_j^b(\theta')})=0
\EN
where $l_{i,j}=\delta_{i,j}-\delta_{i,j+1}-
\delta_{i,j-1}$ and
$K_{1,1}(x)=K_{2,2}(x)
=-\frac{\sqrt{3}}{2cosh(x)-1}$;
$K_{1,2}(x)=K_{2,1}(x)
=-\frac{\sqrt{3}}{2cosh(x)+1}$.
$\nu_i^a= \delta_{i,1} mRcosh(\theta)~(
\lambda<0)$ and $\nu_i^a=\frac{MR}{2}
\delta_{i,1} +\frac{MR}{2}
\delta_{i,p-3}$~$(\lambda>0)$.

The corresponding free-energy $E(R)$
at temperature $T=\frac{1}{R}$ is given
by
\EQ
E(R)= -\frac{1}{2 \pi R}
\sum_{i=1}^{p-3} \sum_{b=1}^{2} \int d \theta
\nu_i^b Ln(1+e^{-\epsilon_i^b(\theta)})
\EN

The $Z_3$-symmetry is expressed
by the kernel $K_{a,b}$ properties and
by the fact that $\epsilon_i^1=
\epsilon_i^2$, $i=1,2,...,p-3$,
is a solution of Eq.(2). For
p=4 we automatically recover the TBA
equations of the 3-state Potts model
perturbed by the thermal field $\phi_{2,1}=\frac{2}{5}$~[14].

The first check consists
of an analysis of Eqs.(2,3) in the
ultraviolet limit,
$R \rightarrow 0$. In this limit the
free-energy $E(R)$ is expected to
behave as $E(R) \simeq -\frac{\pi c}{6 R}$~[15],
where c is the associated
central charge. Using the standard calculations
of the leading behaviour
$R \rightarrow 0$ of the free-energy E(R)~[14,16],
the central charge can be
written in terms of the Rogers-Dilogarithm function L(x) as
\EQ
c_{UV}= 2 \sum_{i=1}^{p-3} \left
[ L(\frac{x_i}{1+x_i})-L(\frac{y_i}{1+y_i})
\right ]
\EN
and
\EQ
L(a)=\frac{3}{\pi^2}
\int_{0}^{a} dx \left [\frac{Ln(x)}{1-x}
+\frac{Ln(1-x)}{x} \right ]
\EN
where the constants
$x_i=e^{-\epsilon_i^1(0)}=e^{-\epsilon_i^2(0)}$
and $y_i=e^{-\epsilon_i^1(\infty)}=e^{-\epsilon_i^2(\infty)}$
are given
by
\EQ
x_a+1= \prod_{i=1}^{a} \frac{sin(\frac{(a+3-i) \pi}{p+1})}
{sin(\frac{i \pi}{p+1})},~~~a=1,2,...,p-3
\EN
and
\EQ
y_1=0;~~y_a+1= \prod_{i=1}^{a-1} \frac{sin(\frac{(a+2-i) \pi}{p})}
{sin(\frac{i \pi}{p})},~~~a=2,3,...,p-3
\EN

Finally, using the Rogers Dilogarithm sum rule
\EQ
\sum_{i=1}^{p-3} L(\frac{x_i}{1+x_i})= \frac{(p-2)(p-3)}{p+1}
\EN
we find the central charge $c_{UV}$ of the minimal $W_p^3$ models
\EQ
c_{UV}=2 \left( 1-\frac{12}{p(p+1)} \right),~~~p=4,5,...
\EN

Turning to the infrared limit
$R \rightarrow \infty$, for the $W_p^{3(-)}$,
it is straightforward to recover the
massive behaviour
$c=0$. In the case of $W_p^{3(+)}$, however,
one may notice that the
pseudoenergies $\epsilon_1$~($\epsilon_{p-3}$)
behave as $\frac{MR}{2}
e^{\theta}$~$(\frac{MR}{2}e^{-\theta})$
in the region $-Ln(MR)<\theta<Ln(MR)$.
Proceeding as in the ultraviolet regime, our final result is
\EQ
c_{IR}=2 \left( 1-\frac{12}{p(p-1)} \right),~~~p=5,6,...
\EN
as it is expected for the RG flow $W_p^{3} \rightarrow W_{p-1}^{3}$.

Now we start to discuss the next to
leading order in R corrections to the
free-energy E(R). Using general
properties of Eq.(2)~[17], it is possible
to show that the functions
$Y_i^a(\theta)=e^{-\epsilon_i^a(\theta)}$ have
a Laurent expansion in terms of the variable
$t=e^{\frac{6 \theta}{p+1}}$. As
a consequence~[17] we expect that the scaling function
$F(r)=\frac{R E(R)}{2 \pi}$~(r=MR or r=mR)
has a series expansion in power
of $R^{\frac{6}{p+1}}$, besides a usual bulk
term proportional to $R^2$.
Moreover, when $p=3i-1,~i=2,3,...$ we
also expect a logarithm divergence
at order $\frac{p+1}{6}$, which gives
a term of $r^2 Ln(r)$ form.

For the sake of clarity, here we
discuss in details the simplest case of
$W_5^{3(\pm)}$ models. From our
considerations above, we have the following
expansion for F(r)
\EQ
F(r)=-\frac{1}{10} -\frac{1}{8 \pi^2} r^2 Ln(r) +\sum_{i=2}^
{\infty}f_i^- r^i,~~~W_5^{3(-)}
\EN
\EQ
F(r)=-\frac{1}{10} -\frac{\sqrt{3}}{24 \pi} r^2
-\frac{1}{8 \pi^2} r^2 Ln(r) +\sum_{i=2}^
{\infty}f_i^+ r^i,~~~W_5^{3(+)}
\EN

The $r^2$ bulk term in the $W_5^{3(+)}$
comes from a simple manipulation
of Eqs.(2,3)~[2,14]. However,
the calculation of the constant $A$ of the
$\frac{A}{{(2 \pi)}^2} r^2 Ln(r)$
term is a bit more involved. Adapting the same method
used by Al. Zamolodchikov for the RSOS model~[1,2], we find that
$A=-\frac{1}{2} \sum_{\stackrel{i=1}{ \{i \neq 3,6,..\}}}^{p-3}
(1+\frac{1}{x_i})= -\frac{3}{p+1}$\footnote
{For $W_p^{3(+)}$ theory, the general form of
the $r^2 Ln(r)$ term is
$(-1)^{\frac{p+1}{3}}
\frac{A}{{(2 \pi)}^2} r^2 Ln(r), p=5,8,...$}. The
coefficients $f_n^+$ and $f_n^-$ can be
calculated solving numerically the
Eqs.(2,3). In table (1) we collected
the first 5 of them. We notice~(within
our numerical precision) that $f_2^+ = f_2^-$, then we
have the same mass scales~(m=M) for both signs of the
perturbation $\lambda$. It is also
important to compare Eqs. (11) and (12)
with the expected results from
the perturbation theory for the
$W_5^{3(\pm)}$ models. Comparing the
logarithm divergence in the
perturbation theory with the term
$- \frac{1}{8 {\pi}^2} r^2 Ln(r)$,
we find the exactly relation
between the mass m~(M) and the
coupling constant $\lambda$
\EQ
m= M =(2 \pi) \lambda
\EN
and from the standard results of
perturbation theory~[1], the coefficients
$f_3^{\pm}$ are given by;
\EQ
f_3^{\pm}= \frac{1}{48}
{(\frac{\lambda}{m})}^3 C_{\Phi_3,\Phi_3,\Phi_3}
{\gamma(\frac{1}{4})}^4
\EN
where $C_{\Phi_3,\Phi_3,\Phi_3}=\frac{1}{3} \gamma(\frac{1}{12})
\gamma(\frac{5}{12}) {\gamma(\frac{3}{4})}^2$\footnote{This structure
constant may be calculated using
the decomposition $c_{W_5^3}=\frac{6}{5}=
\frac{1}{2} +\frac{7}{10}$~[22]}, and
$\gamma(x)=\frac{\Gamma(x)}{\Gamma(1-x)}$. We estimate
$f_3^{\pm}=\mp 3.71575994 E-03$,
\footnote{We have also checked that,
for the next model
$W_6^3$, the coefficients
$f_3^{\pm}$ are in agreement
with the perturbation theory. The corresponding
structure constant may be calculated noticing that
$c_{W_6^3}=\frac{10}{7}$
appears in the superconformal
minimal models $SM_p$, $p=12$.~[23]}
in accordance with the TBA results from Table 1.
Moreover, we observe that the
interchange of signs of form $f_n^+=(-1)^n f_n^-,~n=2,3,...$~(see Table 1)
is in agreement
with what one expects
from the perturbation theory around the positive and negative
signs of $\lambda$.

In order to get some
insight about the next to leading corrections in the
infrared regime, we have
solved numerically the Eqs.(2,3) for large R, namely
in the region $50<R<400$. The first correction is determined to be
$R^{-2}$ with very high precision. This suggests that the infrared behaviour
of the $W_5^{3(+)}$ model may be seen as the $W_4^3$ theory perturbed by
the field $T \bar{T}$~(plus other irrelevant fields)
\EQ
A_{W_5^{3(+)}}= A_{W_4^3} +\int d^2 x T \bar{T}(x)
\EN
where $T (\bar{T})$ is the holomorphic~(anti-holomorphic) component of the
stress-energy tensor of the $W_4^3$ model.

We stress that a similar situation happens in the tricritical Ising model
flowing to the Ising theory~[1,2]. We can check this last proposal as
follows. From perturbation theory~[1,2]
the coefficients $a_1, a_2$ proportional
to the terms $a_1 R^{-2}$ and $a_2 R^{-4}$ satisfy the relation
$\frac{a_1^2}{a_2}=-\frac{c_{IR}}{24}=-\frac{1}{30}$.
Our numerical calculation
predicts $\frac{a_1^2}{a_2}=-0.033332(\pm1)$, in good accordance with
the perturbation theory.

For general $p>5$, the infrared behaviour is mainly governed~(as discussed
before) by the pseudoenergies
$\epsilon_i^1=\epsilon_i^2, i=2,3,...,p-4$. This
changes the periodicity of the function
$Y_i^a(\theta)=e^{-\epsilon_i^a(\theta)}$, and predicts
a new expansion to F(r) in powers of $R^{-\frac{6}{p-1}}$.
This exponent
agrees with the conformal dimension of the
$\phi \left [ \begin{array}{cc}
2 & 1 \\ 2 & 1 \\ \end{array} \right ]$ operator in the $W_{p-1}^3$ minimal
models.
Then, we may argue that this field is one of those that are
responsible\footnote{Such
argument has been originally used by Al.Zamolodchikov
in the case of the usual RSOS models~[1,2]} for the
infrared corrections in the $W_p^{3(+)} $ models.

The generalization of our
results for arbitrary N is as follows. The TBA
equations are written as
\EQ
\epsilon_i^a(\theta) -\nu_i^a(\theta)
+\frac{1}{2 \pi} \sum_{j=1}^{p-N} l_{i,j}
\sum_{b=1}^{N-1} \int_{- \infty}^{\infty}d \theta'
K_{a,b}^{i,j}(\theta-\theta')
Ln(1+e^{-\epsilon_j^b(\theta')})=0
\EN
and the Kernel $K_{a,b}^{i,j}$ is given by
\EQ
K_{a,b}^{i,i}(\theta)=-i\frac{d}{d \theta}Ln(S_{a,b}(\theta));~~~
K_{a,b}^{i,j}(\theta)=i\frac{d}{d \theta}Ln(\tilde {S}_{a,b}(\theta)),~~i \neq
j
\EN
where $S_{a,b}$ is the
two-body S-matrix of the Z(N)-parafermionic field
theory perturbed by the energy operator with anomalous dimension
$\Delta=\frac{2}{N+2}$~[18]. The $\tilde {S}_{a,b}$
has the following expression
\begin{eqnarray}
\tilde {S}_{a,b}(\theta) = \prod_{l=1}^{min(a,b)}
F_{|a-b|+2l-1}(\theta), & &
F(\theta)_m =  \frac{Sh \frac{1}{2}(\theta -\frac{i \pi m}{N})}
{Sh \frac{1}{2}(\theta +\frac{i \pi m}{N})}
\end{eqnarray}

One can notice that the only
poles in the physical strip are due to
the two-body S-matrices $S_{a,b}
$of the Z(N)-model. Then, the functions
$\nu_i^a$ have the form
\begin{eqnarray}
\nu_i^a =  \delta_{i,1} m_a Rcosh(\theta),~
(\lambda<0); & &
\nu_i^a =  \frac{M_a R}{2}
e^{\theta} \delta_{i,1} +\frac{M_a R}{2} e^{-\theta}
\delta_{i,p-N},~(\lambda>0)
\end{eqnarray}
where the masses $m_a$ and $M_a$
are given in terms of the masses ratios
of the Z(N) theory, namely
$m_a=m \frac{sin(\frac{a \pi}{N})}{sin(\frac{\pi}{N})}$
; $M_a=M \frac{sin(\frac{a \pi}{N})}{sin(\frac{\pi}{N})}$,
$a=1,2,...,N-1$. The
free-energy E(R) keeps the same form as Eq.(3), only substituting
$\sum_{i=1}^{p-3}$ by the general form $\sum_{i=1}^{p-N}$.

As in  $N=3$ case, the Z(N)-symmetry of
the pseudoenegies $\epsilon_i^j$~($j=1,2,...,N-1$)
are guaranteed by the
properties of the kernel $K_{a,b}^{i,j}$.
We also have two particular
limits of Eqs.(16,17) that are easy to be taken.
For $N=2$, the Eqs.(16,17) are
the same ones proposed by Al.Zamolodchikov
to describe the minimal models
perturbed by the $\phi_{1,3}$ field.
In the case of arbitrary N and $p=N+1$,
they reproduce the known results for
the Z(N)-models perturbed by the leading
energy operator~[14,18].

Proceeding as in the case $N=3$, we have checked the ultraviolet
and infrared behaviours of Eqs.(16,17). We find that  both
regimes
are in accordance with the central charge of the minimal
models $W_p^{N}$~(p=N+1,N+2,...) related to the $WA_{N-1}$-algebras.

In summary we have proposed and investigated the TBA equations of the
minimal models $W_p^N$ perturbed by the least~($Z_N$ invariant)
primary relevant field
$\Phi_N$. The introduction of an extra degree of freedom, $i.e.$ the
``flavour'' $a=1,2,...,N-1$,
is the basic characteristic of Eqs.[2,16], which allows one to
generalize our results for other symmetries.
In our considerations we  have presented only the final results
without entering into
technical details. Following ref.[19] it is also possible to extend
our results to a general class of coset models based on the SU(N) group.
A detailed
account of the technical part of this letter as well as the generalization
to the SU(N) coset will be published elsewhere~[24].

\section*{Acknowledgements}
It is a pleasure to thank G.Sotkov and M.Stanishkov for discussions on
ref.[22].


\newpage
\centerline{\bf Tables }
Table 1-- The coefficients $f_i^-$, and $f_i^+$ for i=2,3,4,5,6
\begin{center}
\begin{tabular}{|l|c|c|c} \hline
i & $f_i^-$ & $f_i^+$  \\ \hline
2 & $2.92560874(\pm 1)E-02$ & $2.92560875(\pm 1)E-02$ \\ \hline
3 & $ -3.7157599(\pm 1)E-03$ &$3.7157603(\pm 2)E-03$ \\ \hline
4 & $ 1.68716(\pm 2)E-04$ &$1.68712(\pm 2)E-04$ \\ \hline
5 & $ 1.4368(\pm 1)E-04$ &$ -1.4367(\pm 1)E-04$ \\ \hline
6 & $ 1.645(\pm 1)E-05$ &$ 1.642(\pm 1)E-05$ \\ \hline
\end{tabular}
\end{center}

\end{document}